\documentclass[journal]{IEEEtran}
\usepackage{amssymb}

\usepackage{cite}  

\usepackage{amsmath,epsfig}
\usepackage[tight,footnotesize]{subfigure}  
\usepackage{subfigure}

\usepackage{amsfonts}
\usepackage{amssymb}
\usepackage{mathrsfs,amsfonts,dsfont}
\usepackage{bm}

\ifCLASSINFOpdf
\else
\fi
\begin{document}
%
\title{Low Bit-Rate and High Fidelity Reversible Data Hiding}
%
%
%

\author{Xiao Chao~Qu, Suah Kim, Run Cui and Hyoung Joong~Kim \thanks{Xiao Chao Qu is with the Center for Information Security Technologies (CIST), Korea University, Seoul 136171, Korea (e-mail: quxiaochao@gmail.com).}
  \thanks{Suah Kim is with the Center for Information Security Technologies (CIST), Korea University, Seoul 136171, Korea (e-mail: suahn@gmail.com).}
   \thanks{Run Cui is with the Center for Information Security Technologies (CIST), Korea University, Seoul 136171, Korea (e-mail: cuirun@korea.ac.kr).}
 \thanks{Hyoung Joong Kim is with the Center for Information Security Technologies (CIST), Korea University, Seoul 136171, Korea (e-mail: khj-@korea.ac.kr).}}

%
%

\markboth{Journal of \LaTeX\ Class Files,~Vol.~11, No.~4, December~2012}%
{Shell \MakeLowercase{\textit{et al.}}: Bare Demo of IEEEtran.cls for Journals}
%



\maketitle

\begin{abstract}
An accurate predictor is crucial for histogram-shifting (HS) based reversible data hiding methods. The embedding capacity is increased and the embedding distortion is decreased simultaneously if the predictor can generate accurate predictions. In this paper, we propose an accurate linear predictor based on weighted least squares (WLS) estimation. The robustness of WLS helps the proposed predictor generate accurate predictions, especially in complex texture areas of an image, where other predictors usually fail. To further reduce the embedding distortion, we propose a new embedding method called dynamic histogram shifting with pixel selection (DHS-PS) that selects not only the proper histogram bins but also the proper pixel locations to embed the given data. As a result, the proposed method can obtain very high fidelity marked images with low bit-rate data embedded.  The experimental results show that the proposed method outperforms the state-of-the-art low bit-rate reversible data hiding method.
\end{abstract}


%
\IEEEpeerreviewmaketitle

\section{Introduction}
%
%
%
%

\IEEEPARstart{R}eversible data hiding (RDH) is a special data hiding technique that the hidden message can be extracted and  the cover image can be restored. The perfect recovery of the cover image is highly desired in some application scenarios,such as medical or military image processing.

To evaluate the performance of a RDH method, the embedding capacity and the quality of the marked image are the two most important metrics. The embedding capacity tells the amount of data that a RDH method can embed into the cover image and the quality of the marked image measures how much distortion has been induced during embedding the given data. Most existing RDH methods aim to reducing the distortion as much as possible given a certain amount of data.

Histogram shifting (HS) is one of the most popular RDH methods which embeds data by histogram modification. HS first constructs a histogram with some extracted feature where a pair of peak and zero bin is identified. Then, an empty bin is created by shifting all the bins between the peak bin and zero bin towards the zero bin by one. Finally, the data can be embedded into the peak bin. The features used to construct the histogram can be pixel value~\cite{ni2006reversible}, prediction error~\cite{hong2009reversible},~\cite{tsai2009reversible}, interpolation error~\cite{luo2010reversible}, transformed coefficients~\cite{xuan2007reversible} and so on.

Recently, many low bit-rate HS based RDH methods have been proposed which aim to producing high quality marked images. For low bit-rate RDH methods, the reduction of the embedding distortion is more important than the embedding capacity, which can be achieved in many different ways in HS based RDH, including better  feature extraction (usually means better prediction error)~\cite{dragoi},~\cite{luo2011reversible}, pixel selection (or sorting)~\cite{li2011efficient},~\cite{sachnev2009reversible}, histogram bin selection (or dynamic histogram shifting)~\cite{chao2010},~\cite{caciula2014},~\cite{Coatrieux2013},~\cite{hwang2010reversible} and better histogram modification method ~\cite{xiaolong2013},~\cite{bo2013}. Better feature extraction method can construct a histogram with very high peak bin which increases the embedding capacity and decreases the shifting distortion.  Pixel selection chooses pixel positions that the data is embedded with less distortion. Histogram bin selection selects the most proper histogram bin to embed the given data, where the least distortion is introduced. Better histogram modification method decreases the distortion as much as possible by using high dimensional histogram or compensation technique.

In this paper, a low bit-rate and high fidelity reversible data hiding method is proposed. First, we propose an accurate predictor based on weighted least squares (WLS) estimation to generate the prediction error histogram with very high peak bin. Then, we propose a novel dynamic histogram shifting with pixel selection (DHS-PS) method which combines the dynamic histogram shifting and pixel selection together. DHS-PS can find the proper histogram bin and pixel location to embed the given data in a unified framework, and the distortion caused by embedding is significantly reduced. With the proposed WLS predictor and DHS-PS combined together, the proposed method can generate very high fidelity marked image.


The outline of this paper is as follows. The proposed WLS based predictor and DHS-PS are introduced in Section~\ref{PM}. Section~\ref{ER} presents extensive experiments to evaluate the proposed method. Section~\ref{conclusion} provides our conclusion.

\section{Proposed Method}
\label{PM}
\subsection{Weighted Least Squares based Linear Predictor}
\begin{figure}[tb]
\centering
\includegraphics[width=1.5in]{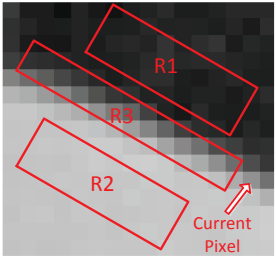}
\caption{An example of the outliers in least squares estimation based predictor. }
\label{outlier}
\end{figure}

Least squares estimation based predictor has been used in~\cite{dragoiLocal}. By updating the estimation weights pixel-by-pixel, the least squares estimation based predictor can adapt to the local image structure and obtain accurate predictions. However, the least squares estimation is easily disturbed by outliers, which leads to incorrect estimated coefficients. As shown in Figure~\ref{outlier}, the current pixel being predicted has similar texture structure with those pixels in region R3 and has different texture structure with those pixels in R1 and R2. However, least squares estimation treats all pixels in region R1, R2 and R3 equally and thus disturbed by those irrelevant pixels in R1 and R2.

Weighted least squares estimation provides the robustness to those outliers by assigning different weights to different pixels according to their relevance to the current pixel. The weights of pixels in R1 and R2 are assigned to relatively small values compared with the weights of pixels in R3. Therefore, WLS estimation emphasizes the minimization of error squares of those pixels in R3. The estimated coefficients precisely reflect the true structure of the current pixel.

Suppose that the current pixel is $y$ and it has $n$  context pixels $[x_1, x_2,..., x_n]$ denoted as $\bm{x}$. The current pixel $y$ can be linearly predicted by its context pixels as

\begin{equation}
\label{linearcoding}
y = \sum_{i=1}^{n}\alpha_nx_n + \beta = \bm{\alpha}\bm{x} + \beta,
\end{equation}
where $\alpha_i$ is the estimated coefficient for $x_i$ and $\bm{\alpha} = [\alpha_1, \alpha_2, ..., \alpha_n]$. $\beta$ is the coding error with small value.

To estimate $\bm{\alpha}$, $m$ relevant pixels are collected into the training set $S$, where each training sample is a pair of one pixel and its context pixels. All those pixels' context pixels are organized into a matrix $\bm{X} \in \mathds{R}^{m \times n}$ as

\[
\bm{X}=
  \begin{bmatrix}
  x_1^1 & x_2^1 & \cdots & x_n^1\\
  x_1^2 & x_2^2 & \cdots & x_n^2\\
  \vdots & \vdots & \ddots & \vdots\\
  x_1^m & x_2^m & \cdots & x_n^m

   \end{bmatrix}
\]

and all the pixels are grouped into a vector $\bm{Y} \in \mathds{R}^{m \times 1}$ as
\[
\bm{Y^T}=
  \begin{bmatrix}
  y^1 & y^2 & \cdots & y^m\\

   \end{bmatrix}
\]

For least squares estimation, the estimated coefficient $\bm{\alpha}$ should minimize the square errors as $||\bm{Y} - \bm{X}\bm{\alpha}||^2_2$, where $|| ||^2$ is  square of the $L_2$ norm of a vector. Weighted least square estimation incorporates a weight matrix $\bm{W} \in \mathds{R}^{m \times m}$ into the estimation process. Assume $w^i$ is assigned to the training sample $(y^i,\bm{x}^i)$ in the training set $S$, the $\bm{W}$ is as follows

\[
\bm{W}=
  \begin{bmatrix}
  w^1 & 0 & \cdots & 0\\
  0& w^2 & \cdots & 0\\
  \vdots & \vdots & \ddots & 0\\
  0 & \cdots & \cdots & w^m

   \end{bmatrix}
\]

WLS estimation minimizes the square errors as $||\bm{W}(\bm{Y} - \bm{X}\bm{\alpha})||^2_2$. The solution to the above minimization problem can be obtained as

\begin{equation}
\label{solution}
\bm{\alpha} = (X^TWX)^{-1}X^TWY.
\end{equation}

The weight $w^i$ in $\bm{W}$ is designed to reflect the image structure relevance between the $i$-th training pixel $y^i$ and the current pixel $y$. In the extracting process of RDH methods, $y$ is unknown when doing the pixel prediction, therefore, the value of $y$ can not be used to calculate $w^i$. As a result, $w^i$ is calculated by using the context pixels of $y$ and $y^i$ as

\begin{equation}
\label{calW}
w^i = \frac{1}{||\bm{x} - \bm{x^i}||^2_2 + \gamma},
\end{equation}
$\gamma$ is a small value to prevent from the dividing by zero problem. As can be seen, $w_i$ is small when the square difference between $\bm{x}$ and $\bm{x_i}$ is large and $w_i$ is large when the square difference  between $\bm{x}$ and $\bm{x_i}$ is small. Because the square difference between two context pixel vectors reflects the local image structure between two pixels, the value of $w_i$ thus reflects the structure relevance between $y$ and $y^i$.

\begin{figure}[tb]
\centering
\includegraphics[width=1.8in]{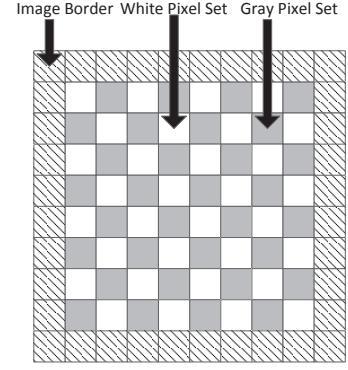}
\caption{The cover image is divided into three parts: image border, white pixel set and gray pixel set. }
\label{imageSplit}
\end{figure}

The cover image is divided into three parts as shown in Figure~\ref{imageSplit}. The image border is not used to embed data and will not be predicted. The White pixel set and gray pixel set are used to embed data and the white pixel set is first used. The pixel prediction in the  two stage embedding scheme takes the advantage of full context prediction and usually produces better prediction results.

\begin{figure}[tb]
\centering
\includegraphics[width=3in]{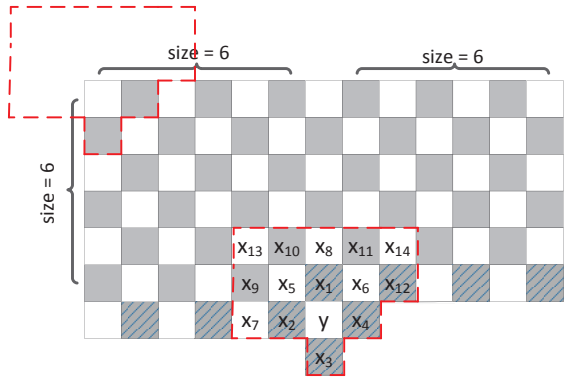}
\caption{The white pixel x is predicted by its contexts from $x_1$ to $x_{14}$ in the red dashed region, and the training set size is controlled by the size parameter. The red dashed region in the up-left corner indicates the range of the pixels involved in the proposed WLS estimation.}
\label{pixelPrediction}
\end{figure}

In the following, we use pixels in the white set as examples to show the detailed prediction process which is same for the gray pixel set. For each pixel $y$ to be predicted, the context pixels $\bm{x}$ are defined as shown in Figure~\ref{pixelPrediction}. The fourteen context pixels can be defined in other ways and the the number of context pixels can be different. All the pixels except y and those pixels with diagonal lines are included in the training set $S$. The white pixels before $y$ are already recovered when predicting $y$ in the extracting process, so that they can be used in predicting $y$. The pixels with diagonal lines can not be used because some context pixels are not accessible when predicting $y$ in the decoder side. In summary, the training set $S$ are same for the embedding process and extracting process to make sure $y$ has same prediction. The overall size of the training set $S$ is controlled by the size parameter as shown in Figure~\ref{pixelPrediction}. After constructing the training set $S$, the prediction of $y$ can be obtained using the proposed WLS estimation process.

\subsection{Dynamic Histogram Shifting with Pixel Selection}

Dynamic histogram shifting and pixel selection both try to reduce the distortion caused by embedding the given data.

\begin{figure}[tbp]
  \centering
     \subfigure[Dynamic histogram shifting ]{
        \includegraphics[width = 1 in]{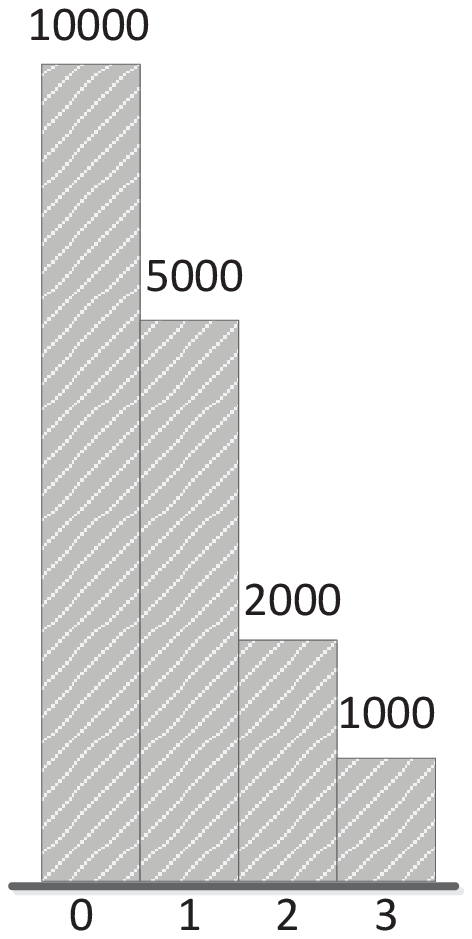}}
    \hspace{0.1in}
     \subfigure[DHS-PS]{
        \includegraphics[width = 1 in]{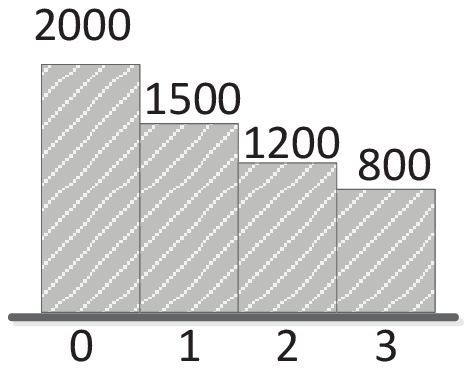}}
    \hspace{0.1in}
    \caption{The comparison between dynamic histogram shifting and DHS-PS.}
    \label{histogramBin}
\end{figure}

Dynamic histogram shifting reduces the distortion in a global manner by selecting the best histogram bin to embed data. In this way, some histogram bins can avoid modification as shown in Figure~\ref{histogramBin}(a). For example, when the payload is $1,000$ bits, the histogram bin $3$ can be selected to embed the payload, and other histogram bins do not need to be modified. The distortion is greatly reduced compared with normal histogram shifting methods which will use the histogram bin $0$ to embed data. However, the histogram is constructed using the whole cover image which usually does not have  satisfactory shape for the given payload. For example, when the payload size is $1,001$, histogram bin $3$ can not provide enough embedding capacity and histogram bin $2$ is thus used to embed data. Histogram $3$ will be shifted to right to create an empty bin which causes large distortion.

Pixel selection reduces distortion in a local manner. Usually, it estimates the local smoothness value of each pixel and embeds data only into those pixels in smooth image region. In this way, pixel selection avoids pixel modifications in complex image regions where it is difficult to embed data. However, the accurate estimation of the smoothness value is not easy by itself, so that pixel selection may choose the inappropriate pixels to use.

We notice that the drawbacks of dynamic histogram shifting and pixel selection can be mitigated by combining them together. The proposed dynamic histogram shifting with pixel selection (DHS-PS) first separates the cover image into smooth image part and complex image part by using pixel selection. Then, for the smooth image part, DHS-PS selects the proper histogram bin to embed data by using dynamic histogram shifting. Compared with dynamic histogram shifting, DHS-PS includes a local operation which selects part of the cover image instead of the whole cover image to reconstruct the histogram. Compared with pixel selection, DHS-PS includes a global operation to select the best histogram bin to embed data. The advantage of DHS-PS is shown in Figure~\ref{histogramBin}(b). The histogram generated by DHS-PS has lower peak bin (due to that part of the cover image is used) than that of dynamic histogram shifting, however, less distortion is caused compared with dynamic histogram shifting when the payload size is $1,001$. The bin $2$ with the height of $1,200$ is used given the payload size of $1,001$. In summary, DHS-PS can generate more proper histograms for a given payload. Histograms with different combinations of bins can be obtained by using different pixel selection thresholds. The smoothness value used can be calculated based on the local neighboring pixel differences as in~\cite{bo2013} or local neighboring pixel prediction errors as in~\cite{qu2014}.

Given a specific payload, DHS-PS thus  needs to search the best pixel selection threshold and histogram bin to use. The exhaustive search of the combinations of pixel selection threshold and histogram bin is very time-consuming or even prohibitive. A greedy algorithm can be used as follows.
\begin{enumerate}
  \item Search pixel selection threshold from a small value to a predefined big value. Construct a histogram using the pixels with smoothness value smaller than the current pixel selection threshold.
  \item Search two proper histogram bins the same as in~\cite{chao2010}.
  \item Embed the payload with the current pixel selection threshold and histogram bin value. The embedding is same as that in~\cite{chao2010}.
  \item Stop when the PSNR value decreases for the first time. Otherwise increase the pixel selection threshold value by 1 and go to step 1.
\end{enumerate}

\subsection{Embedding Process and Extracting Process}

The embedding process is as follows.
\begin{enumerate}
  \item Preprocess the cover image $I$ into $I_1$ to avoid overflow and underflow problem. All pixels with the value of $0$ are modified into $1$ and all pixels with the value of $255$ are modified into $254$. A location map is used to record all the modifications  and compressed using arithmetic coding. The proposed algorithm modifies the pixel value at most by $1$, so that $I_1$ will not have overflow and underflow problem.
  \item $I_1$ is divided into image border (includes the first and last rows and the first and last columns), the white pixel set and the gray pixel set. Divide the payload into two halves and embeds the first half payload and the first half compressed location map into the white pixel set to obtain $I_2$ .
  \item Embed the second half of the payload and the second half compressed location map into $I_2$ to get $I_3$.
  \item Embed some overhead information into the image border: 1) Embeds the pixel selection threshold, histogram bin used for white pixel set and  the  compressed location map size into the first and last rows. 2) Embeds the pixel selection threshold, histogram bin used for gray pixel set and  the  compressed location map size into the first and last columns.
\end{enumerate}

The extracting process is as follows.
\begin{enumerate}
  \item Extract the overhead information in the image border.
  \item Extract data from the gray pixel set and recover the original pixel values.
  \item Extract data from the white pixel set and recover the original pixel values.
  \item Decompress the compressed location map and recover the original cover image.
\end{enumerate}

\section{Experiment}
\label{ER}
In this section, we will validate the superior performance of the proposed WLS estimation predictor and DHS-PS. All the test images (except the Barbara) used in the following experiments are from the SIPI image database\footnote[1]{http://sipi.usc.edu/database.} and are eight-bit gray-scale images with the size $512 \times 512$.

\begin{figure}[tbp]
  \centering
 \subfigure[Lena]{
        \includegraphics[width = 0.5 in]{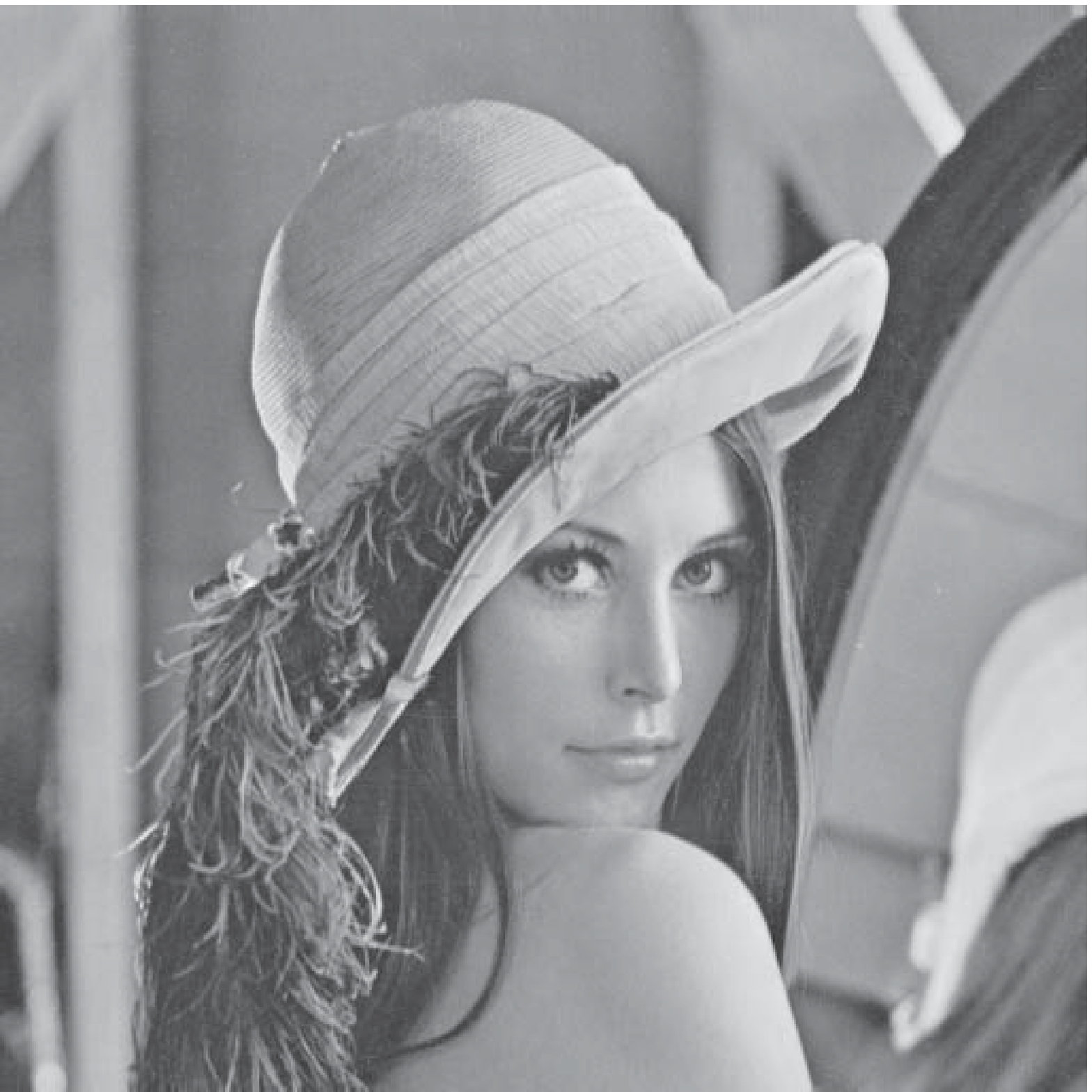}}
    \subfigure[F16]{
        \includegraphics[width = 0.5 in]{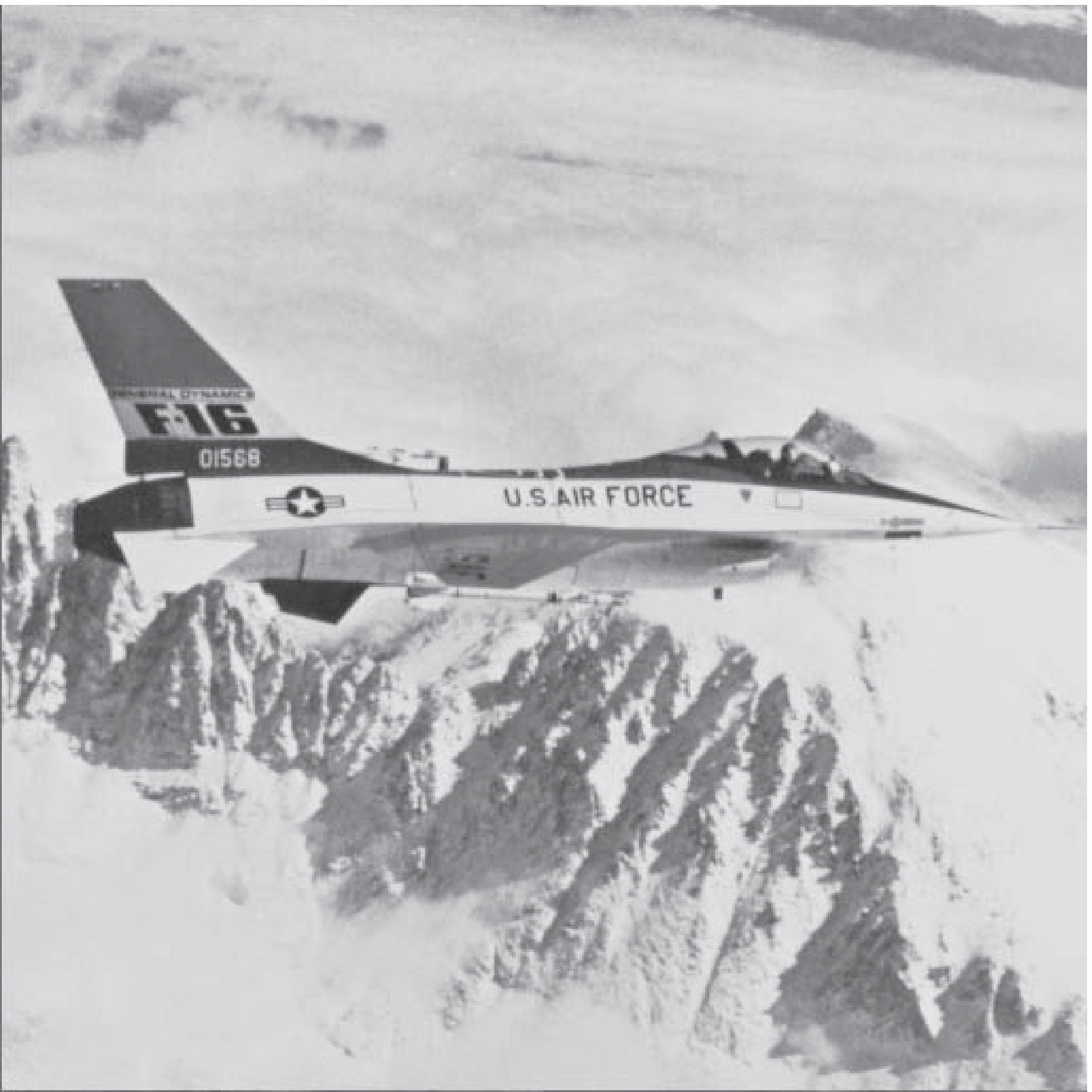}}
  \subfigure[Baboon]{
    \includegraphics[width = 0.5 in]{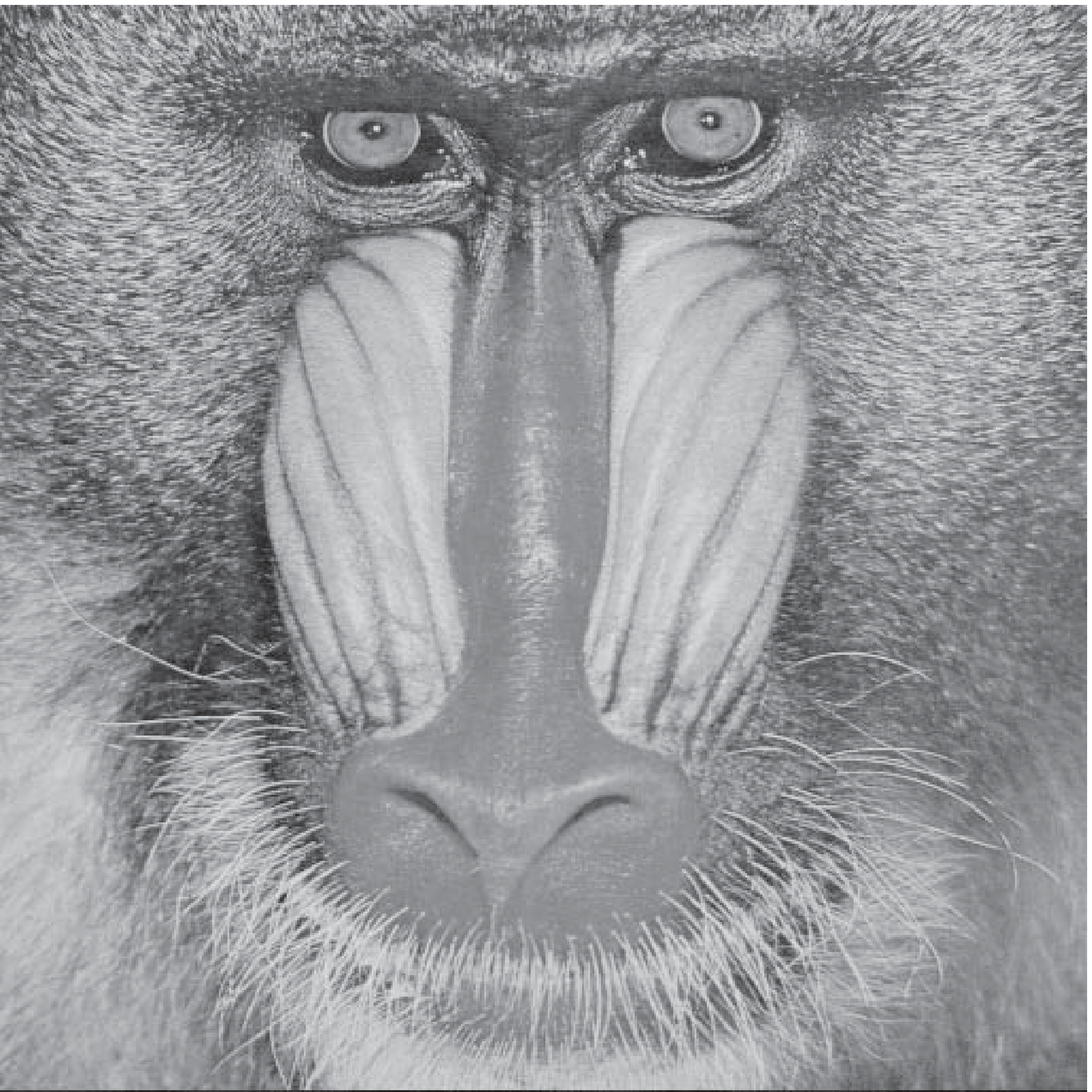}}
    \subfigure[Barbara]{
        \includegraphics[width = 0.5 in]{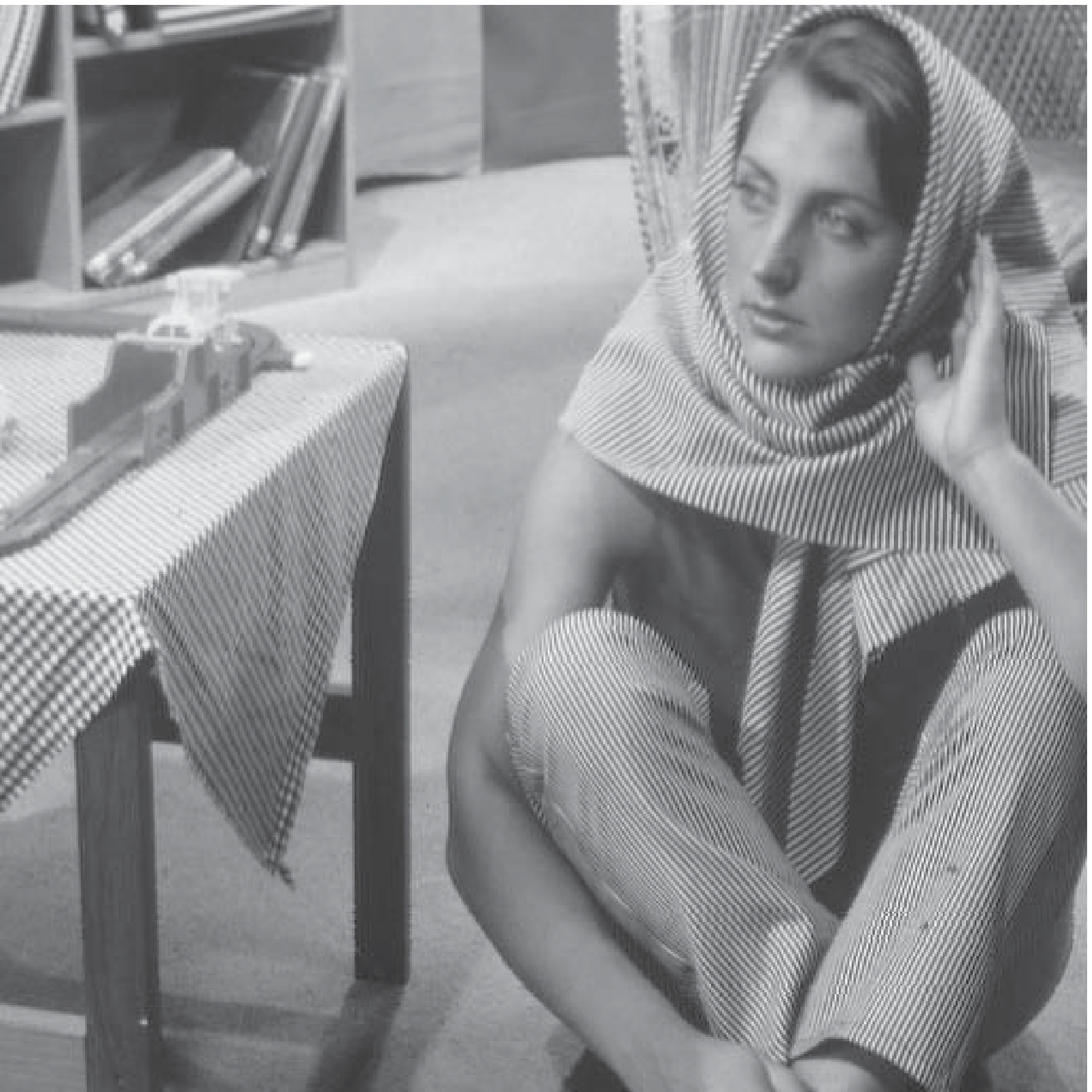}}
  \subfigure[Boat]{
    \includegraphics[width = 0.5 in]{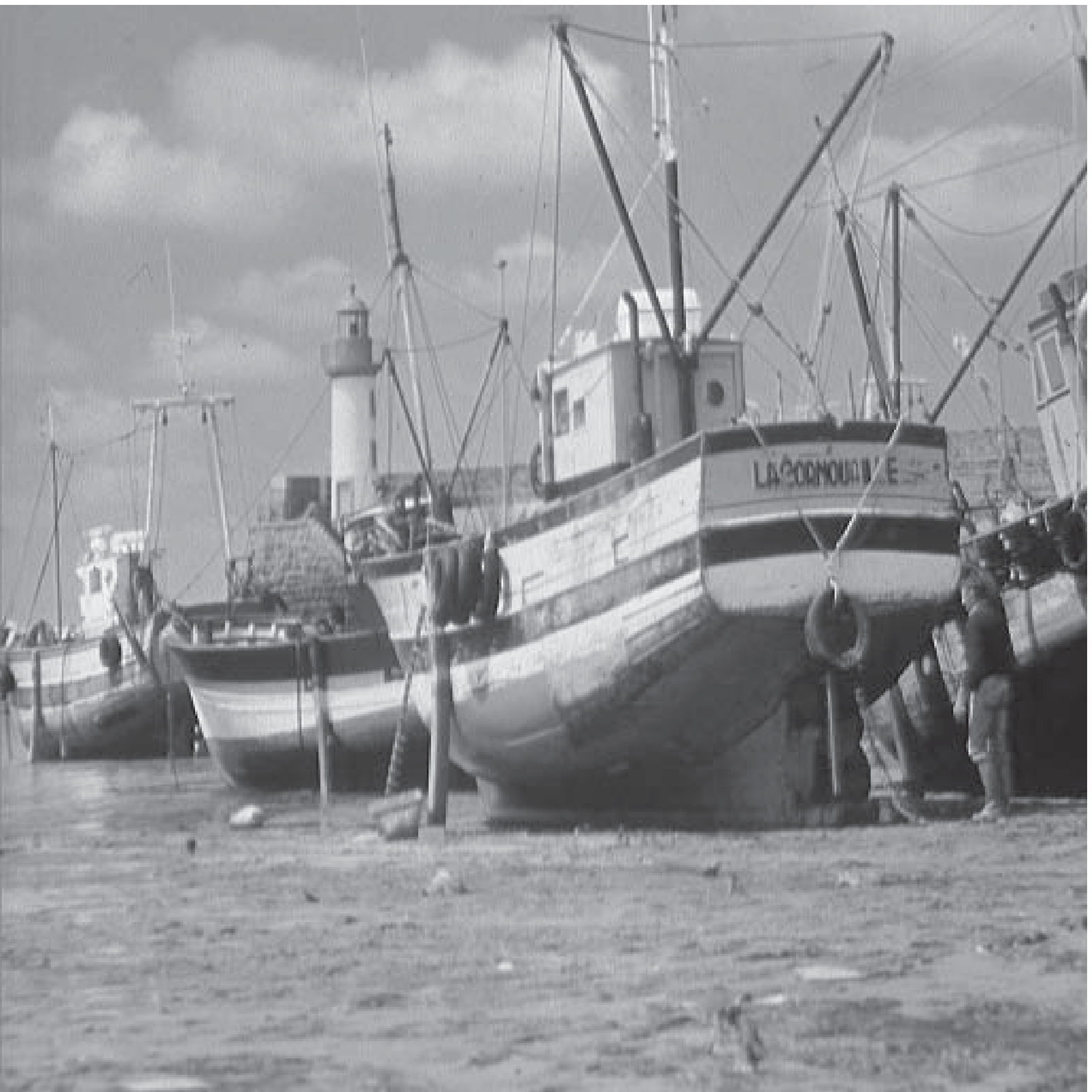}}
    \subfigure[Peppers]{
        \includegraphics[width = 0.5 in]{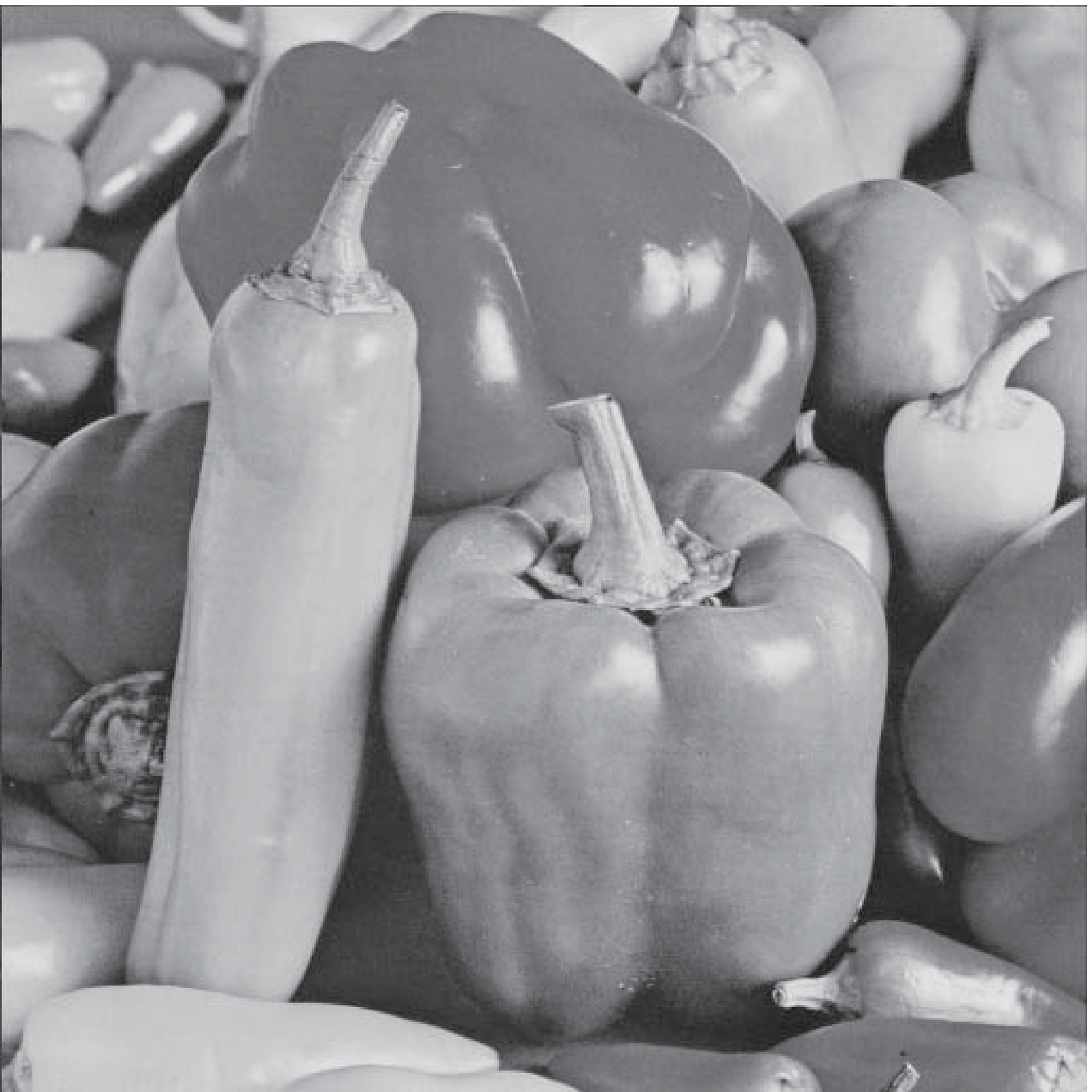}}
    \subfigure[Sailboat]{
        \includegraphics[width = 0.5 in]{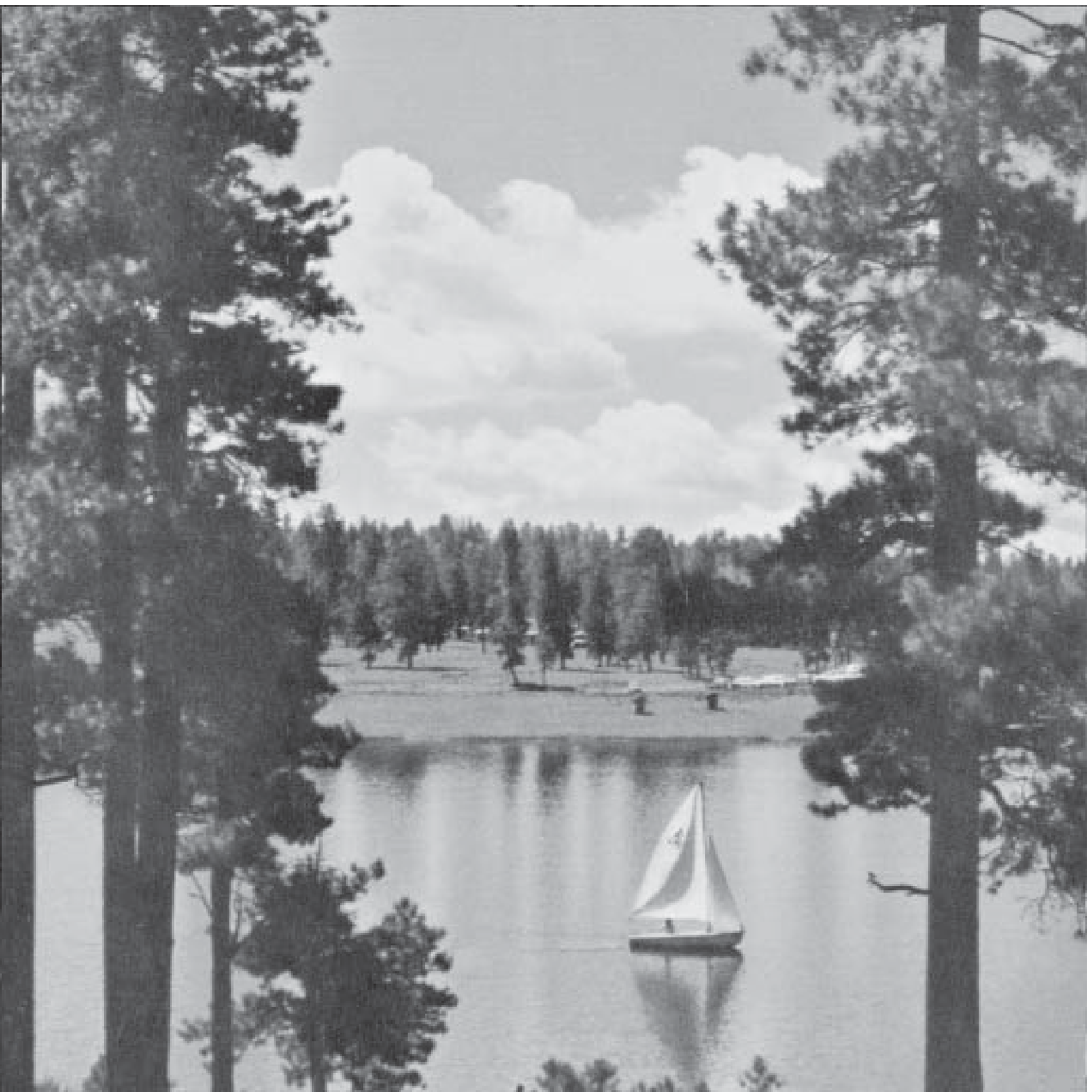}}
  \subfigure[Elaine]{
    \includegraphics[width = 0.5 in]{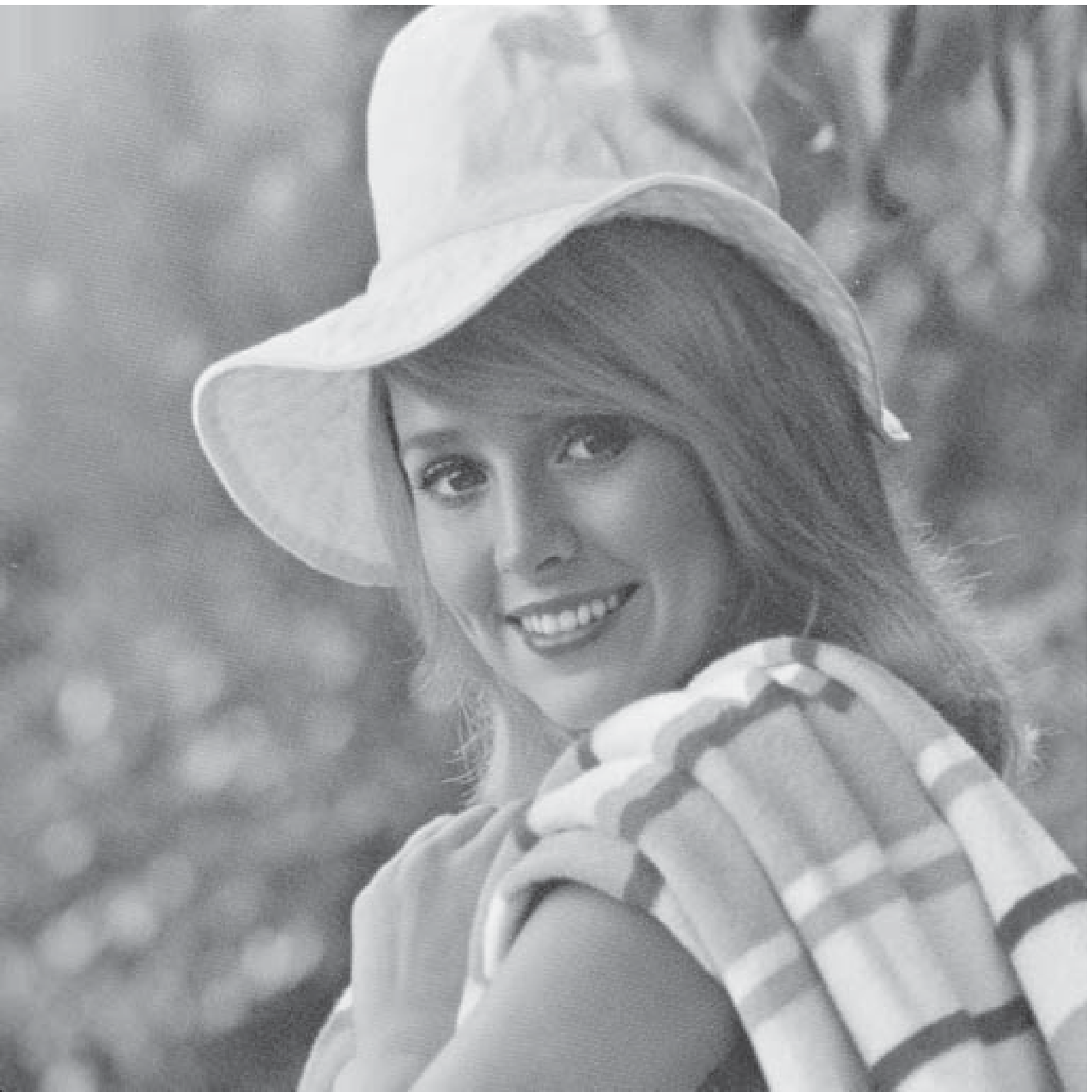}}
    \caption{The test images of SIPI image data set}
    \label{testImage}
\end{figure}
First, we use the entropy as the metric to evaluate the prediction performance of the proposed WLS estimation predictor and compare its entropy with that of several other widely used predictors, where MED is used in~\cite{hong2009reversible} and GAP is used in~\cite{xiaolong2013}. Assume $p_i$ is the occurrence probability of  histogram bin $i$, the entropy is defined as

\begin{equation}
\text{entropy} = -\sum_{i = 1}^N p_i log_2(p_i),
\end{equation}
where $N$ is the total number of histogram bins. The entropy is shown in Table~\ref{entropy}. The number of context pixels and the size of training set are both set to be $10$ because we found $10$ is a proper value for most testing images.  As can be seen, LS and WLS have much lower entropy than that of MED and GAP, and WLS has the lowest entropy.
\begin{table}[ptb]
\caption{Comparisons in  terms of entropy for different predictors.}
\vspace{0.5em}\centering
\begin{tabular}{ c |c |  c | c | c  }
\hline
\hline
Image &  MED &  GAP &  LS &  WLS  \\
\hline
Lena & 4.55  & 4.39   & 3.87  & \textbf{3.86} \\
\hline
F16&4.18& 4.12 & 3.60 & \textbf{3.58} \\
\hline
Baboon&6.27&6.21&5.61& \textbf{5.60}\\
\hline
Barbara&5.48&5.38&4.07&\textbf{3.99}\\
\hline
Boat&5.10&4.97&4.29&\textbf{4.24}\\
\hline
Peppers&4.94&4.72&4.33&\textbf{4.31}\\
\hline
Elaine&5.34&5.15&4.65&\textbf{4.62}\\
\hline
Sailboat&5.38&5.25&4.83&\textbf{4.81}\\
\hline
Average&5.15&5.02&4.40&\textbf{4.37}\\
\hline
\hline
\end{tabular}
\label{entropy}
\end{table}

\begin{figure*}[tb]
  \centering
  \subfigure[Lena]{
        \includegraphics[width = 2.5 in]{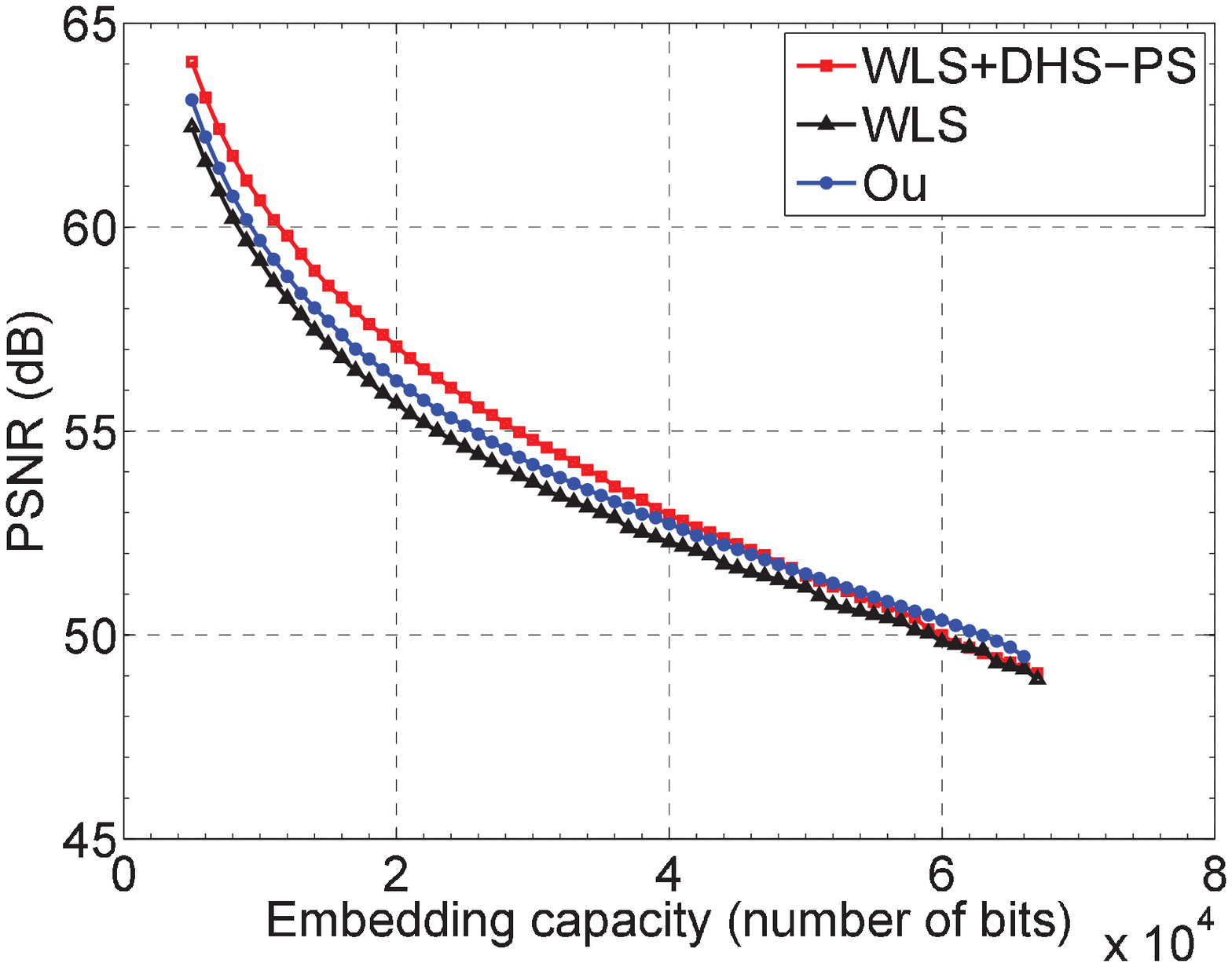}}
    \subfigure[F16]{
        \includegraphics[width = 2.5 in]{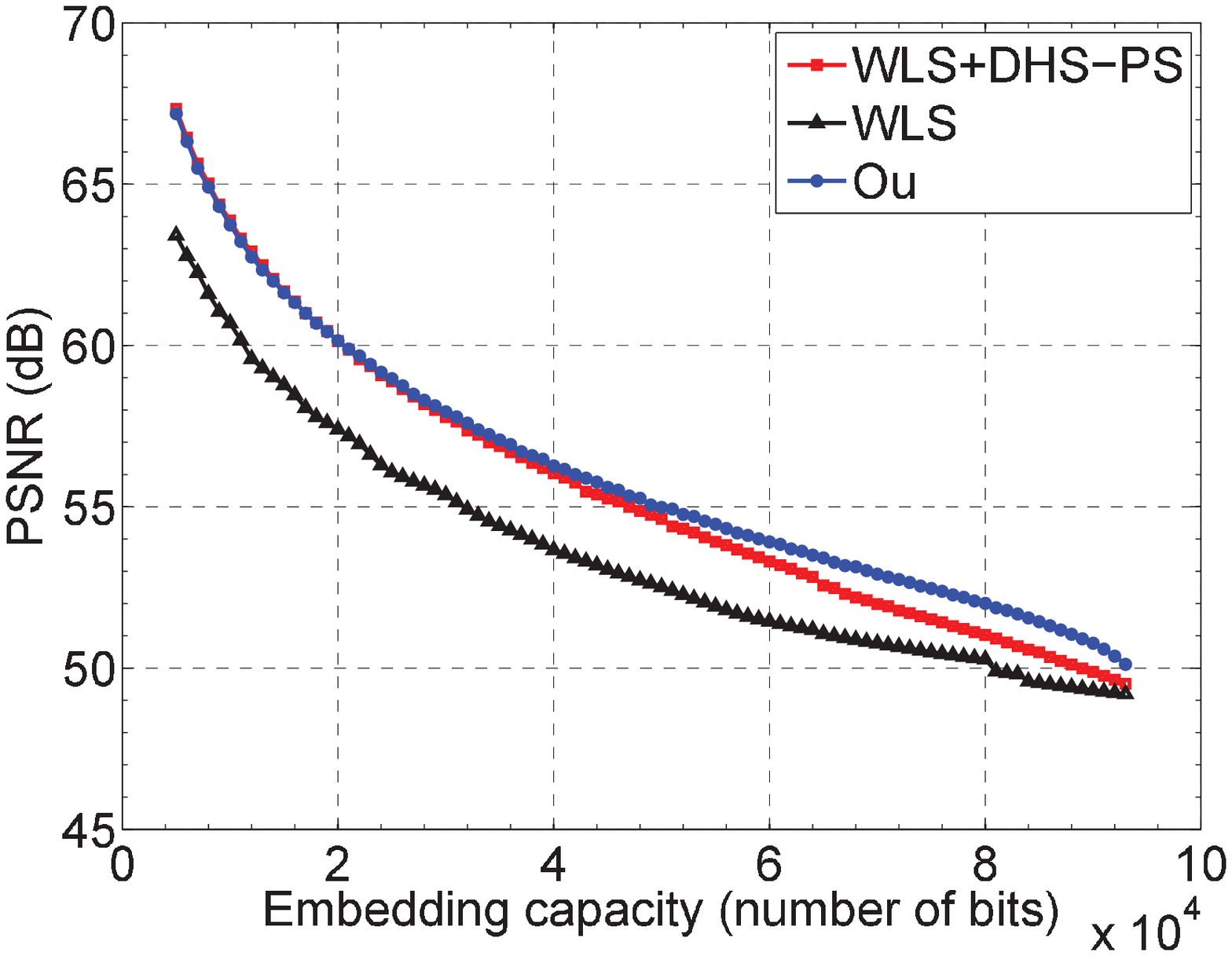}}
  \subfigure[Baboon]{
    \includegraphics[width = 2.5 in]{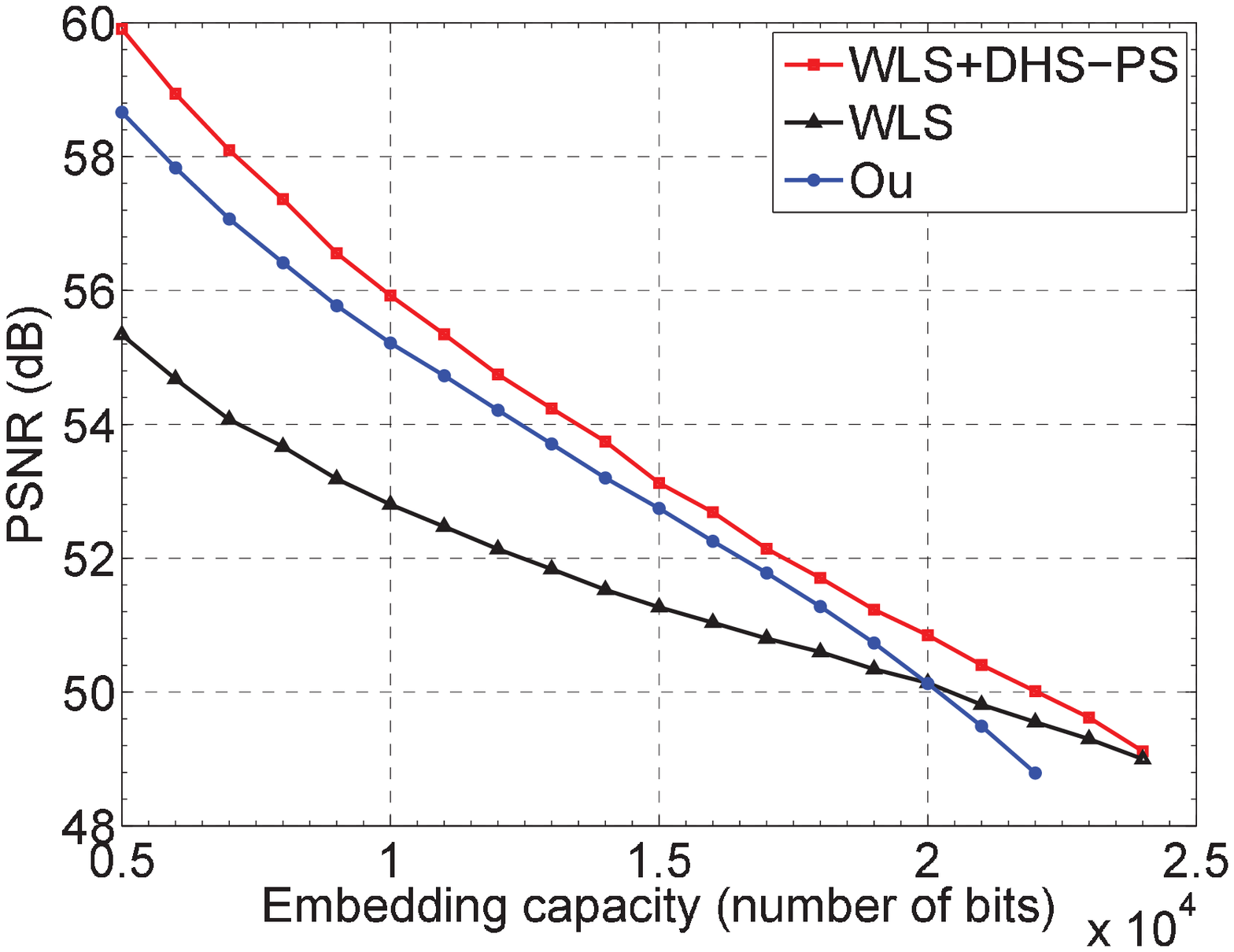}}
    \subfigure[Barbara]{
        \includegraphics[width = 2.5 in]{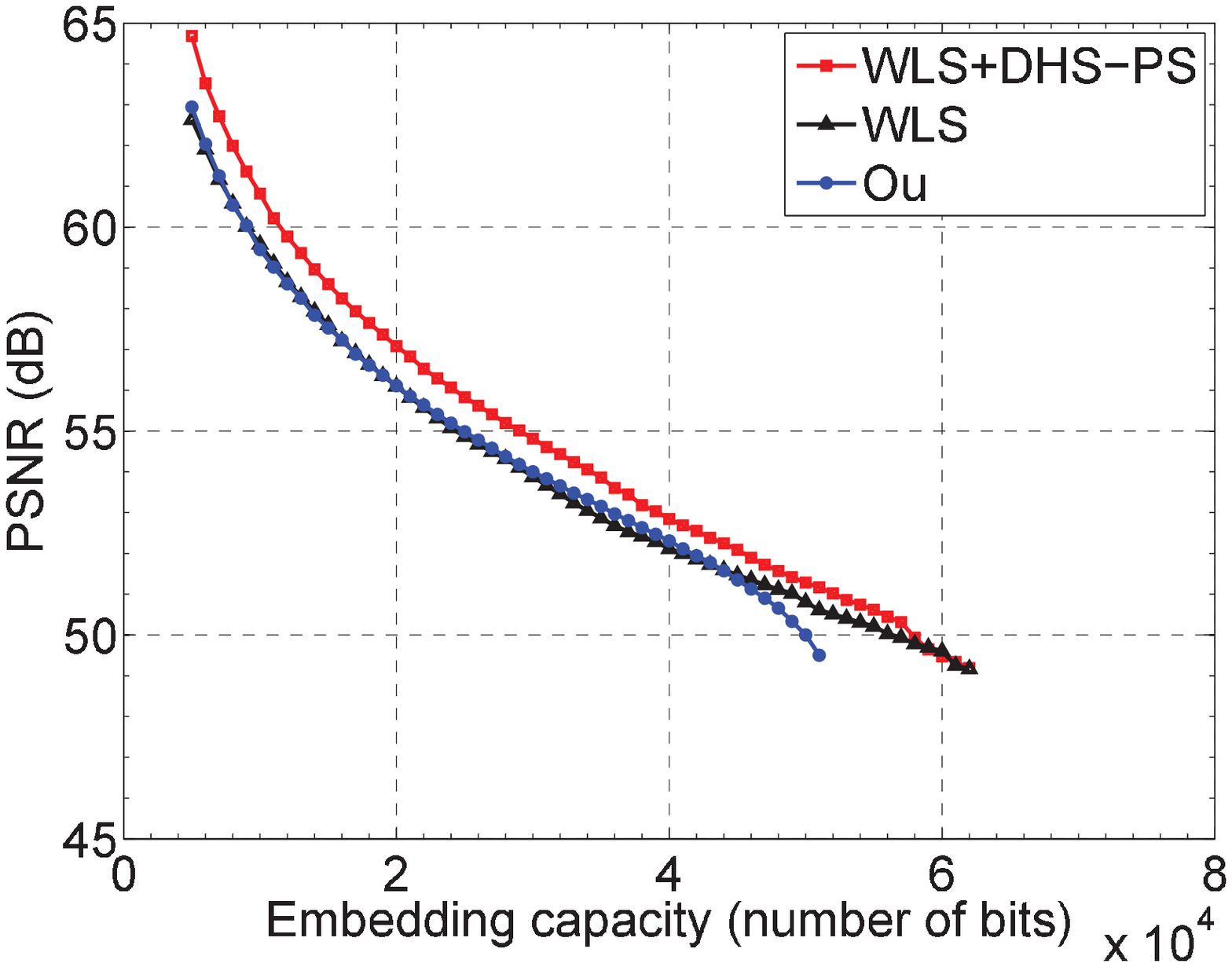}}
    \subfigure[Boat]{
        \includegraphics[width = 2.5 in]{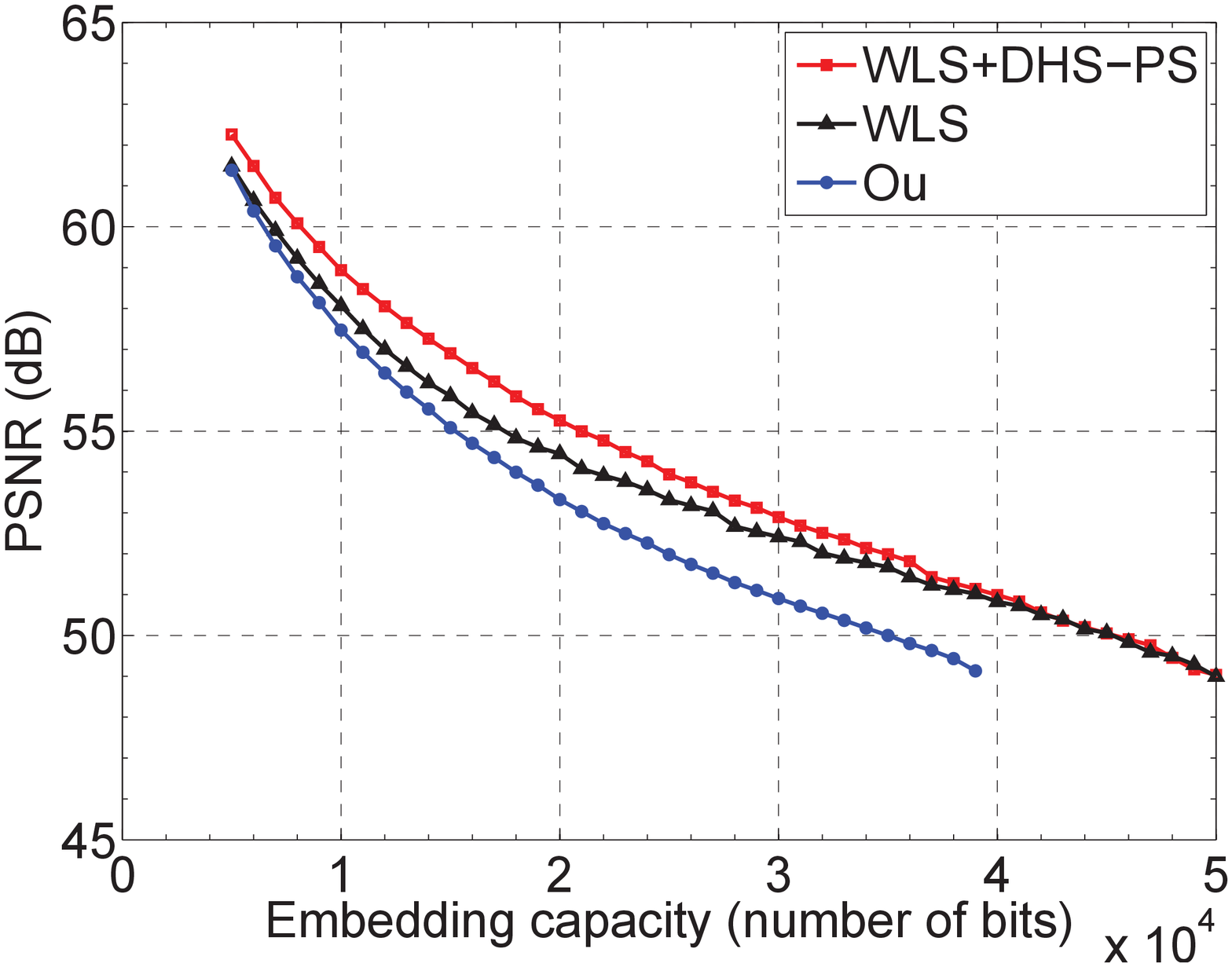}}
    \subfigure[Peppers]{
        \includegraphics[width = 2.5 in]{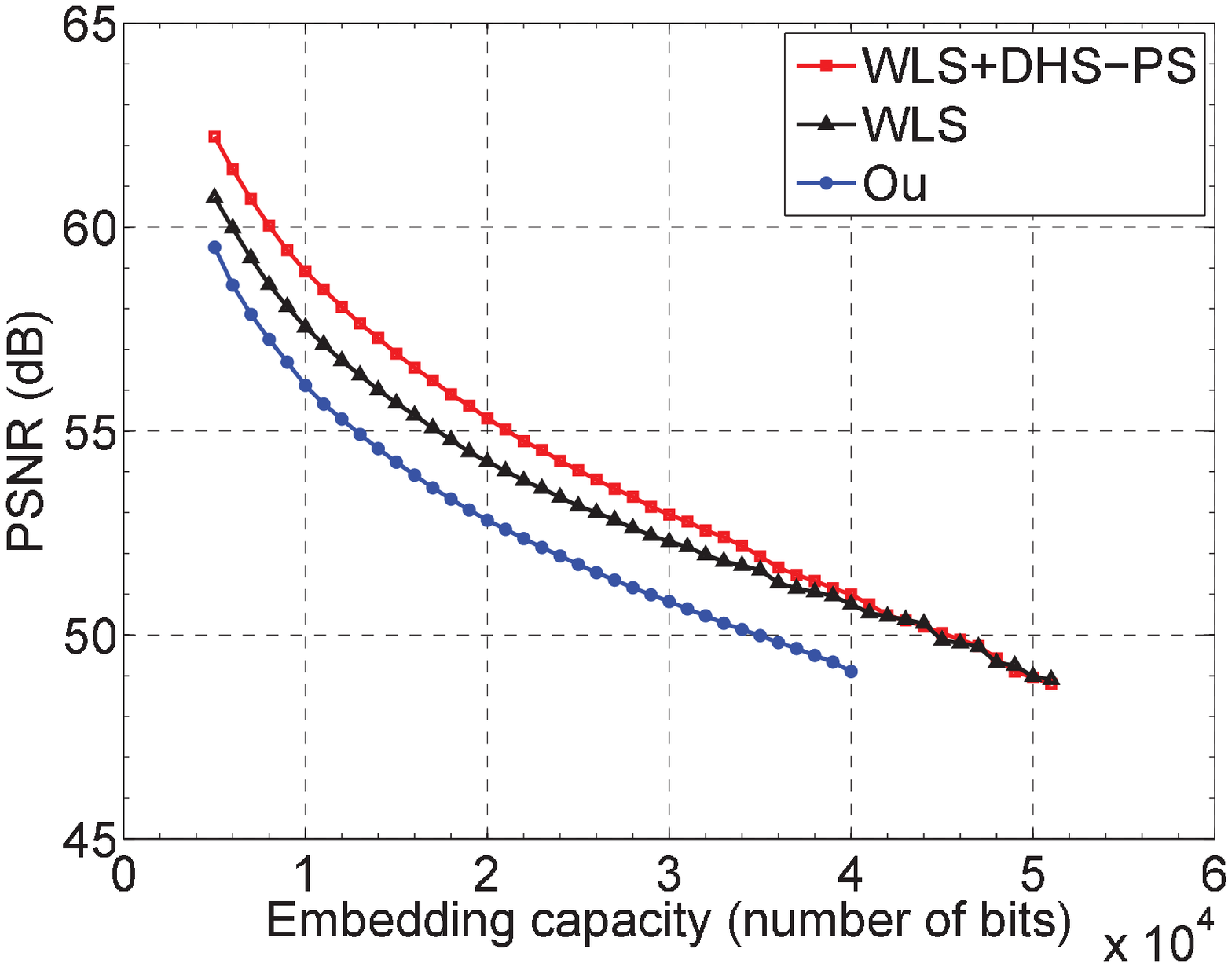}}
  \subfigure[Elaine]{
    \includegraphics[width = 2.5 in]{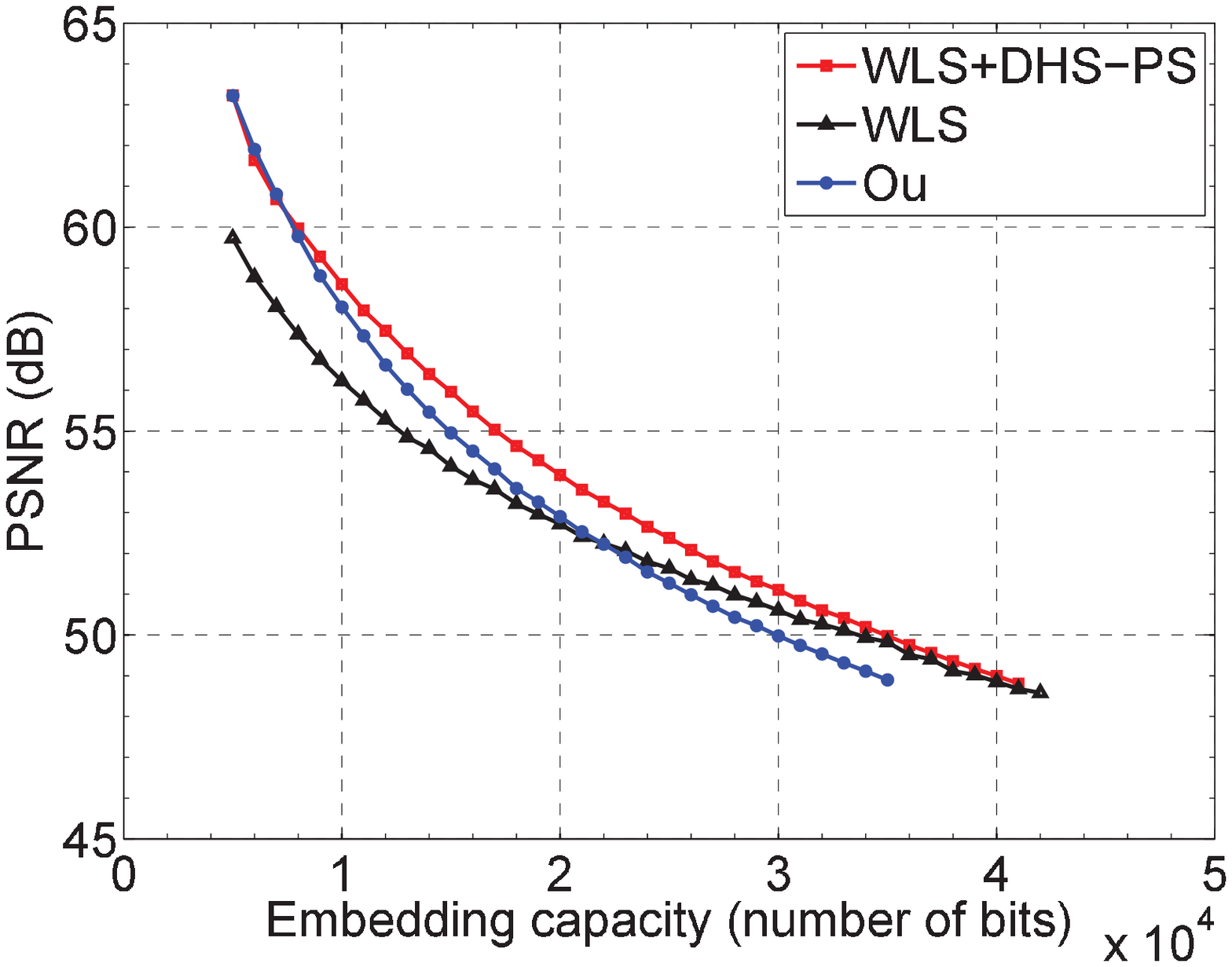}}
    \subfigure[Sailboat]{
        \includegraphics[width = 2.5 in]{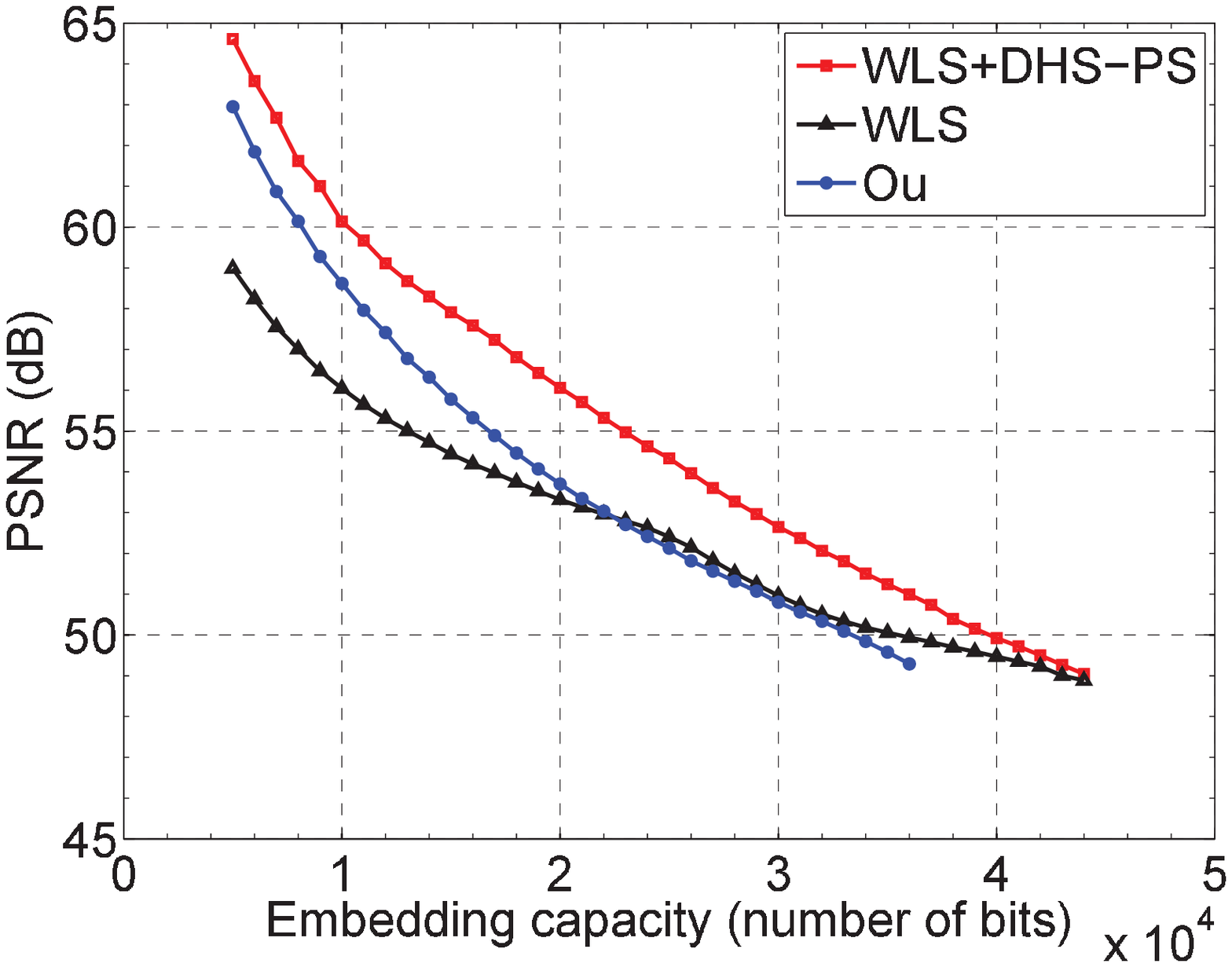}}
    \caption{The performance comparison  of WLS+DHS-PS, WLS and Ou~\cite{bo2013} for  the test images of SIPI image data set.}
    \label{performance}
\end{figure*}

Next, we perform the following experiment to show the effectiveness of the proposed DHS-PS. The embedding capacity and the peak signal-to-noise ratio (PSNR) value are used as the evaluation metrics. The experiment results are shown in Figure~\ref{performance}. Three algorithms are compared with each other: the first algorithm uses both WLS and DHS-PS, the second algorithm uses only WLS and the third algorithm is proposed by Ou~\cite{bo2013} which is the best low-bit rate RDH method. As can be seen, the first algorithm performs much better than the second algorithm with small payload size. For example, the PSNR value of the Baboon image for the first algorithm  and the second algorithm are  $55.92$ dB  and  $52.80$ dB, respectively. The proposed DHS-PS helps the first algorithm increase the PSNR value by $3.12$ dB. However, with the increase of the payload size, the PSNR value of the first algorithm and the second algorithm will converge to similar values. The reason is that the proposed DHS-PS has to select the peak bins to use when the payload size is large. As a result, there is no difference with or without DHS-PS when the payload size is large.

Compared with Ou~\cite{bo2013}, it can be seen that the first algorithm performs better for most of the images. When the image is very smooth (e.g. F16), Ou~\cite{bo2013} performs better due to its two dimensional histogram shifting scheme which reduces distortion significantly. However, for images with complex textures (e.g. Baboon), the proposed first algorithm performs better. The combination of WLS and DHS-PS achieves the state-of-the-art performance as far as we know.

\section{Conclusion}
\label{conclusion}
In this paper, we propose an accurate weighted least squares (WLS) based linear predictor and a novel dynamic histogram shifting with pixel selection (DHS-PS) for high fidelity and low bit-rate reversible data hiding. Extensive experiment verifies that the proposed method can obtain very high quality marked images and outperforms the state-of-the-art method.

%
\bibliographystyle{abbrv}
\bibliography{mybib}  

\end{document}